\documentclass[twocolumn,showpacs,preprintnumbers,amsmath,amssymb]{revtex4-1}

\usepackage{graphicx}
\usepackage{bm}

\begin{document}

\title{First order phase transition in the few-body XY-models with surface fields}

\author{Xintian Wu}

\affiliation{Department of Physics, Beijing Normal University,
Beijing, 100875, China}

\date{\today}

\begin{abstract}
We investigate the one-dimensional finite-size XY model with opposing surface fields in the X direction. Exact solutions are obtained for the two-site and three-site models, while numerical methods are employed for models with more than three sites. Remarkably, first-order quantum phase transitions are observed in this system. At the phase transition point, the energy gap closes linearly, and the magnetization at each site undergoes a discontinuous jump. Additionally, we identify a $Z_2$ symmetry that accompanies the phase transition and its associated symmetry change. Notably, the first-order phase transition in finite-size systems does not exhibit the conventional finite-size rounding effect. On the contrary, there exists a counterintuitive finite-size effect: the amplitude of the jump in magnetization at each site decreases as the lattice size increases. Interestingly, lattices with an even number of sites share a common phase boundary, while lattices with an odd number of sites have a distinct phase boundary.
\end{abstract}

\maketitle
\section{Introduction}

Quantum phase transitions (QPTs) have garnered significant interest in recent decades, driven by experiments on various new materials, particularly topological materials in condensed matter and novel quantum systems provided by cold atoms \cite{sachdev,vojta,chiu,islam}.

On one hand, there has been considerable attention on few-body quantum phase transitions (QPTs) discovered in the Quantum Rabi Model (QRM) \cite{hwang,ying,liu,cong,cong1,liu1,liu2,ashhab}. Traditionally, phase transitions are perceived to occur in the thermodynamic limit in condensed matter systems. However, the QRM, composed of a single qubit or spin-half system coupled with a light field or a bosonic mode, exhibits notable few-body QPTs.

On the other hand, extensive studies have focused on the finite-size scaling in first-order quantum phase transitions in recent years \cite{vicari-prl,vicari-1,vicari-2,vicari-3,vicari-4,vicari,wu1,wu2}. In typical first-order quantum phase transitions, the transition is rounded due to finite-size effects.

This paper explores a model that exhibits a few-body phase transition, which can be considered "fewer" than the QRM. Unlike the QRM, where the spin couples with an infinite-dimensional bosonic mode, our model has a finite-dimensional Hilbert space that can be quite small. It can be exactly solved for small lattices and numerically solved for larger lattices. Furthermore, the first-order phase transition in our model is not affected by finite-size effects. On the contrary, the smaller the lattice size, the more pronounced the singularity of the phase transition. In the thermodynamic limit, the phase transition disappears.

It is widely established, following Yang and Lee's seminal works \cite{yang1,yang2,yang3}, that thermal phase transitions only occur in the thermodynamic limit. Rigorous proofs indicate that thermal phase transitions cannot happen in one-dimensional infinite classical and quantum spin lattices with finite-range interactions \cite{ruelle,araki}. However, in our study, we specifically address quantum phase transitions on finite-size lattices, for which no similar proof exists. While counterintuitive, we demonstrate explicit non-analyticity in the exact solutions of the two-site and three-site models. Numerical solutions reveal non-analytic behaviors for systems with more than three sites $N>3$. We observe that the non-analyticity weakens as the lattice size increases and vanishes in the thermodynamic limit, aligning with the rigorous proof of no phase transitions in one-dimensional infinite lattices.

As is commonly known, first-order phase transitions are typically accompanied by a symmetry change. We identify a $Z_2$ symmetry in the Hamiltonian, which we refer to as parity symmetry. The exact solutions of the two-site and three-site models explicitly demonstrate different parities for different phases.

The paper is organized as follows: In Section II, we define the model and present the phase diagrams. Section III presents the results for lattices with an even number of sites, while Section IV focuses on the results for lattices with an odd number of sites. In Section V, we discuss the Jordan-Wigner transformation for this model and its dynamics. Finally, Section VI provides a summary.

\section{The model and the phase diagrams}

We consider a lattice-based spin model comprising $N$ sites. The model's Hamiltonian is given by:
\begin{equation}
H=-\frac{1}{4}\sum_{i=1}^{N-1}(\sigma_i^x \sigma_{i+1}^x+\sigma_i^y \sigma_{i+1}^y)-\frac{h_L}{2}\sigma_1^x+\frac{h_R}{2}\sigma_N^x
\end{equation}
Here, $h_L$ and $h_R$  represent positive values, and it is noteworthy that the surface fields, namely 
$h_L$ and $-h_R$ exhibit opposite orientations. This model can be regarded as an isotropic XY-model characterized by opposing surface fields in the X direction.

This Hamiltonian possesses a conserved quantity that commutes with the generator:
\begin{equation}
\hat{P}_x=\prod_{i=1}^{N}  \sigma_i^x .
\end{equation}
It is evident that $\hat{P}_x^2=1$, thus for any eigenstate $|\Psi\rangle$ of the Hamiltonian , we have $\hat{P}_x|\Psi\rangle=P_x|\Psi\rangle=\pm |\Psi\rangle$. This represents a $Z_2$ symmetry, often referred to as parity symmetry. In the basis in which $\sigma_i^z$ is diagonal, i.e. $\sigma_i^z |s_i^z\rangle=s_i^z  |s_i^z\rangle$ where $s_i^z=\pm 1$, $\hat{P}_x |s_1^z,s_2^z,\cdots,s_N^z\rangle=|-s_1^z,-s_2^z,\cdots,-s_N^z\rangle$. To simplify the discussion, we define an eigenstate to have even parity if $\hat{P}_x|\Psi\rangle= |\Psi\rangle$ and odd parity if $\hat{P}_x|\Psi\rangle=-|\Psi\rangle$.

In the subsequent discussion, we will predominantly employ the representation where $\sigma_i^x$ is diagonal, specifically given by $\sigma_i^x=\left ( \begin{array}{cc}
            1 &  0 \\
            0 &  -1
           \end{array} \right )$.  The states will be denoted as $|s_1^x,s_2^x,\cdots,s_N^x\rangle$, 
where $s_i^x$ corresponds to the eigenvalues of $\sigma_i^x$. Consequently, the boundary field terms and the coupling term  $\sum_i\sigma_i^x \sigma_{i+1}^x$ become diagonal in this representation.  Furthermore, in this representation, we find that $\sigma_i^y=\left ( \begin{array}{cc}
            0 &  1 \\
            1 &  0
           \end{array} \right )$. Thus  $\sigma_i^y \sigma_{i+1}^y|\cdots, s_i^x,s_{i+1}^x,\cdots \rangle=|\cdots, -s_i^x,-s_{i+1}^x,\cdots \rangle$. In other words,  $\sigma_i^y \sigma_{i+1}^y$ reverses the spins of the $i$th and $(i+1)$th sites in any state while preserving the parity. This straightforwardly explains the preservation of parity in this model.

Due to the conservation of parity and the possibility to classify states based on their parity, the Hilbert space is divided into two distinct parts: one with even parity and the other with odd parity. In general, we can express this as follows:
\begin{equation}
H=\hat{H}^{(+)}+\hat{H}^{(-)}
\label{eq:oddeven}
\end{equation}
with
\begin{equation}
\hat{H}^{(+)}=\left ( \begin{array}{cc}
            H^{(+)} &  0 \\
            0         &  0
           \end{array} \right )
~~~,\hat{H}^{(-)}=\left ( \begin{array}{cc}
            0      &  0 \\
            0      &H^{(-)}
           \end{array} \right )
\end{equation}
where $H^{(+)} $ corresponds to the matrix defined in the Hilbert space with even parity, while $H^{(-)}$ corresponds to the matrix defined in the Hilbert space with odd parity. Two explicit examples that demonstrate this separation are provided by the exact solutions for the two-site and three-site models described below. Importantly, we observe that:
\begin{equation}
[\hat{H}^{(+)}, \hat{H}^{(-)} ]=0
\label{eq:commute}
\end{equation}
which implies that $\hat{H}^{(+)}$ and $\hat{H}^{(-)}$  commute to each other. 

In the context of the first-order quantum phase transition, a significant phenomenon known as level-crossing occurs. As mentioned earlier, the Hamiltonian can be decomposed into two mutually commuting parts, each operating on different subspaces \cite{hami}. As the system parameters undergo variation, the ground state has the potential to transition from being an eigenstate of $H^{(+)}$ to that of $H^{(-)}$, or vice versa. This occurrence aligns with one of the possibilities of level-crossing elaborated upon by Sachdev at the beginning of his book, "Quantum Phase Transition" \cite{sachdev}.

By evaluating the energy gap and observing the magnetization profile, we have discovered the presence of a first-order phase transition that occurs as the parameters $h_L$ and $h_R$  undergo variation. To visually represent these findings, we have constructed phase diagrams. Interestingly, we observe that lattices with an even number of sites exhibit a consistent phase diagram, while lattices with an odd number of sites demonstrate a distinct phase diagram, as illustrated in Figure 1(c) and 1(d).

\begin{figure}
\includegraphics[width=0.5\textwidth]{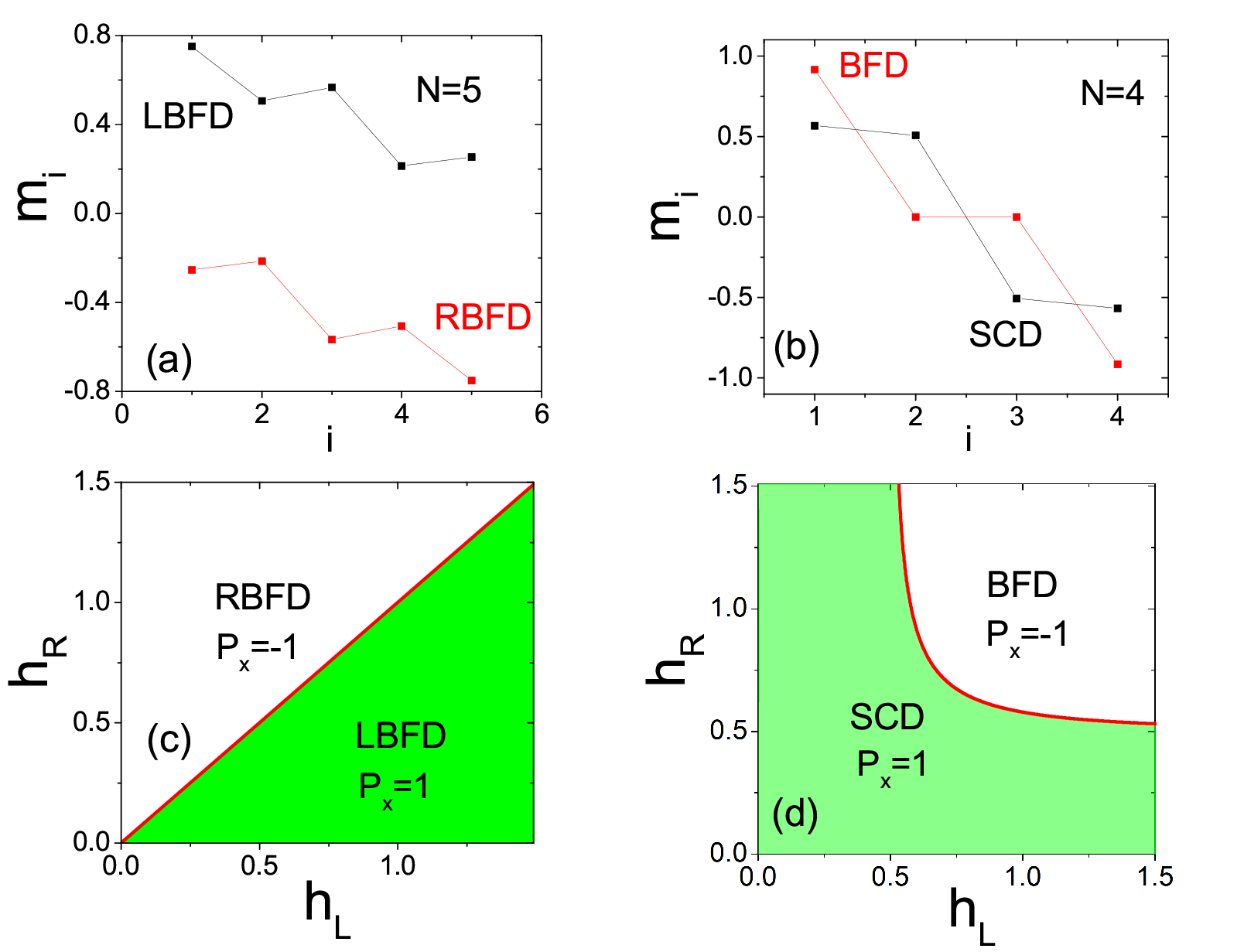}
\caption{(a) Magnetization profiles for $N=5$ in two different phases: LBFD phase and RBFD phase. (b) Magnetization profiles for $N=4$ in two different phases: BFD phase and SCD phase. (c) The phase diagram for lattices with  odd sites . The separatrix, indicated by a solid red line, separates the two phases by the condition $h_L=h_R$. The parity $P_x$ is obtained from Eq. (\ref{eq:parity2}). (d) The phase diagram for all the lattices with even sites.  The phase boundary, represented by a solid red line, is determined by Eq. (\ref{eq:pt}). The parity $P_x$ is obtained from Eq. (\ref{eq:parity1}). }

\end{figure}

In the exact solution for  $N=3$, we have demonstrated that when  $h_L>h_R$ , the ground state exhibits even parity, denoted by $P_x=1$, as depicted in Figure 1(c). Conversely, when  $h_L<h_R$ , the ground state possesses odd parity, indicated by  $P_x=-1$. We have named the phase corresponding to $h_L>h_R$  as the Left Boundary Field Dominating (LBFD) phase, and the phase for $h_L<h_R$  as the Right Boundary Field Dominating (RBFD) phase. It is worth noting that for systems with an odd number of sites greater than three, they share the same parity behavior as the 
$N=3$ system. In general, phase transitions often occur between states with distinct symmetries, and this particular phase transition adheres to that principle.

In the figure 1(a), we present the magnetization profile ($m_i=\langle 0|\sigma_i^x |0\rangle$, where $|0\rangle$ denotes the ground state) of a representative system with an odd number of spins (N=5) in the two distinct phases. The LBFD phase is depicted by a black line and square in Fig. 1(a) , while the RBFD phase is presented in red line and square,  while the RBFD phase is represented by a red line and square in Figure 1(a). In the phase diagram shown in Figure 1(c), the phase boundary occurs when $h_L=h_R$ . When $h_L>h_R$, the total magnetization displayed in Figure 1(a) is positive. In this scenario, the sign of the total magnetization aligns with the left boundary field, thus establishing the dominance of the Left-Boundary-Field phase. Conversely, when $h_L<h_R$, the total magnetization becomes negative. In this case, the sign of the total magnetization aligns with the right boundary field, indicating the dominance of the Right-Boundary-Field phase.

The phase transition in systems with an even number of spins differs from that in systems with an odd number of spins, as illustrated in Figure 1(b) and 1(d). For such systems, there exist two distinct phases: the BFD phase (represented by the green region in Figure 1(d)) and the SCD phase (depicted by the white region in Figure 1(d)). These phases can be understood by considering two limiting cases. In the SCD phase, when the boundary fields 
$h_L$  and $h_R$  approach zero, the influence of the boundary field terms becomes negligible compared to the couplings between spins. Hence, we refer to this phase as the Spin-Coupling Dominating (SCD) phase. Conversely, in the BFD phase (the white region), as $h_L$  and $h_R$  approach infinity, the boundary fields clearly dominate over other factors. Therefore, we refer to this phase as the Boundary Fields Dominating (BFD) phase.

In the exact solution of $N=2$, it is demonstrated that  the ground state exhibits  even parity with $P_x=1$ in the SCD phase, while in the BFD phase the ground state exhibits odd parity with $P_x=-1$ as shown in Fig. 1(d). For systems with an even number of sites, they are expected to display the same parity as the $N=2$ system.

Figure 1(b) presents the magnetization profile for a representative system with an even number of spins (in this case $N=4$). The red line and squares correspond to the BFD phase, while the black line and squares represent the SCD phase. The BFD data is obtained with double precision, precisely above the transition point at $h_L=h_R=\frac{\sqrt{2}}{2}+10^{-12}$. Conversely, the SCD data is obtained just below the transition point at $h_L=h_R=\frac{\sqrt{2}}{2}-10^{-12}$.  In both phases, the total magnetization is zero due to the anti-symmetric nature of the magnetization profiles. However, the magnetization values for individual spins differ above and below the transition point. Within the BFD phase, the absolute values of the magnetization for the two boundary spins exceed those in the SCD phase. This discrepancy highlights the substantial difference in boundary field energy between the two phases.

\section{Even sites lattices}

\subsection{Exact solution for $N=2$}

To commence, let us tackle the most elementary scenario, in which we address the case of a lattice containing two sites, denoted by $N=2$. It can be diagonalized directly. This particular configuration readily lends itself to diagonalization. By virtue of the conserved parity $P_x$, we can classify states based on their parity characteristics. We now employ a basis in which the operator  $\sigma_i^x$ assumes a diagonal form. Introducing the notation $s_i^x=+1,-1$ to represent the eigenvalues associated with two distinct states, we denote these states as $|1\rangle_i,|-1\rangle_i$ each pertaining to its respective index $i=1,2$. In the context of a two-spin system, we find ourselves confronted with four unique states. The two states $|\psi_1\rangle =|1,1\rangle$, $|\psi_2\rangle=|-1,-1\rangle$ possess even parity, whereas the remaining two states  $|\psi_3 \rangle = |1,-1\rangle$, $|\psi_4\rangle=|-1,1\rangle$ exhibit odd parity. With this foundation in place, we are equipped to present the expression for the Hamiltonian as follows:
\begin{equation}
H=\left ( \begin{array}{cc}
            H^{(+)} &  0 \\
            0      &H^{(-)}
           \end{array} \right )
\end{equation}
where $H^{(+)}$ where $H^{(+)} $ corresponds to the matrix defined in the Hilbert space with even parity, 
\begin{equation}
H^{(+)}=-\frac{1}{4}\left ( \begin{array}{cc}
            2(h_L-h_R)+1 &  1 \\
            1     & -2(h_L-h_R)+1 
           \end{array} \right ),
\end{equation}
and $H^{(-)}$ corresponds to the matrix defined in the Hilbert space with odd parity.
\begin{equation}
H^{(-)}=-\frac{1}{4}\left ( \begin{array}{cc}
            2(h_L+h_R)-1 &  1 \\
            1     & -2(h_L+h_R)-1 
           \end{array} \right )
\end{equation}
Its first two eigenvalues are given by
\begin{eqnarray}
E_1 & = &\frac{1}{4}[-1-\sqrt{1+4(h_L-h_R)^2}], \nonumber \\
E_2 & = &\frac{1}{4}[1-\sqrt{1+4(h_L+h_R)^2}].
\label{eq:e-two}
\end{eqnarray}
$E_1$ is obtained from $H^{(+)}$ and the corresponding state has even parity and  $E_1$ is obtained from $H^{(-)}$ and the corresponding state has odd parity .

In the scenario where $\frac{1}{h_L^2}+\frac{1}{h_R^2}>4$, the eigen-energy $E_1$ assumes the lowest value, thereby establishing the first state as the ground state with an even parity phase, denoted by $P_x=1$.  This outcome arises from the diagonalization of the Hamiltonian $H^{(+)}$. Conversely, when  $\frac{1}{h_L^2}+\frac{1}{h_R^2}<4$,  the eigen-energy $E_2$ becomes the smallest, resulting in the second state becoming the ground state with an odd parity phase, given by $P_x=-1$. Here $E_2$ is obtained from the diagonalization of the Hamiltonian $H^{(-)}$. The determination of the ground state's parity is based on these results. Therefore we have
\begin{equation}
P_x=\left \{ \begin{array}{cc}
            1, ~~~&  \frac{1}{h_L^2}+\frac{1}{h_R^2}>4 \\
            -1,~~~     & \frac{1}{h_L^2}+\frac{1}{h_R^2}<4 
           \end{array} \right.
\label{eq:parity1}
\end{equation}
At the point where the equation
\begin{equation}
\frac{1}{h_L^2}+\frac{1}{h_R^2}=4
\label{eq:pt}
\end{equation}
is satisfied, a level crossing takes place.

In the given scenario, where $h_L=h_R=h$, the phase transition point is located at $ h_L=h_R=\frac{\sqrt{2}}{2}$. By substituting this value into Eq. (\ref{eq:pt}), we can determine the energies $E_1$ and $E_2$ as follows $E_1=-\frac{1}{2}$ and $E_2=\frac{1}{4}-\sqrt{1+(4h)^2}$. The energy gap, denoted as $\Delta$,can be calculated as the absolute difference between $E_1$ and $E_2$. However, it exhibits a non-analytic behavior at the phase transition point. Specifically, we have:
\begin{equation}
\Delta=|E_1-E_2|=\left \{ \begin{array}{cc}
                  \frac{3}{4}-\frac{1}{4}\sqrt{1+(4h)^2}, & h<\frac{\sqrt{2}}{2} \\
                  \frac{1}{4}\sqrt{1+(4h)^2}-\frac{3}{4}, & h>\frac{\sqrt{2}}{2}
                \end{array} \right.
\label{eq:delta}
\end{equation}
At the phase transition point $h=\frac{\sqrt{2}}{2}$, the energy gap closes. It is important to note that this expression is singular because the derivative  $\frac{\partial \Delta }{\partial h}$  is discontinuous at $h=\frac{\sqrt{2}}{2}$. Specifically, we have:
\begin{equation}
\frac{\partial \Delta }{\partial h} =\left \{ \begin{array}{cc}
 -\frac{2\sqrt{2}}{3}, & h=\frac{\sqrt{2}}{2}-0^+ \\
  \frac{2\sqrt{2}}{3}, & h=\frac{\sqrt{2}}{2}+0^+ .
 \end{array} \right.
\label{eq:delta-ab}
\end{equation}
The energy gap curve exhibits a very sharp cusp at the transition point $h=\frac{\sqrt{2}}{2}$. This non-analytic behavior indicates the presence of a phase transition.

The corresponding first two eigen-states for $h_L=h_R=h$ can be expressed as:
\begin{eqnarray}
|\Psi_1\rangle & = & \frac{1}{\sqrt{2}}(|1,1\rangle+|-1,-1\rangle) \nonumber \\
|\Psi_2\rangle & = & a(\frac{1}{4}|1,-1\rangle+(\sqrt{h^2+(\frac{1}{4})^2}-h)|-1,1\rangle).
\label{eq:psi12}
\end{eqnarray}
Here $a^{-2}=2\sqrt{h^2+(\frac{1}{4})^2}(\sqrt{h^2+(\frac{1}{4})^2}-h)$. The average of $\sigma_1^x,\sigma_2^x$  for these two states are 
\begin{eqnarray}
\langle\Psi_1|\sigma_1^x|\Psi_1\rangle & = & \langle\Psi_1|\sigma_2^x|\Psi_1\rangle=0, \nonumber \\
\langle\Psi_2|\sigma_1^x|\Psi_2\rangle & = & -\langle\Psi_1|\sigma_2^x|\Psi_1\rangle=-\frac{4E_2 a^2}{h}.
\label{eq:m1m2}
\end{eqnarray}
When  {\small $h<\frac{\sqrt{2}}{2}$}, the ground state is{\small  $|\Psi_1\rangle $}  in this region,  {\small $m_1=\langle\Psi_1|\sigma_1^x|\Psi_1\rangle$} . On the other hand,  when $h>\frac{\sqrt{2}}{2}$, the ground state is {\small $|\Psi_2\rangle$  in this region,   {\small $m_1=\langle\Psi_2|\sigma_1^x|\Psi_2\rangle$}. Thus, the surface magnetization in the X-direction on the left side can be defined as:
\begin{equation}
m_1\equiv \langle 0|\sigma_1^x|0\rangle =\left \{
\begin{array}{cc}
            0 ,~~~~     & h<\frac{\sqrt{2}}{2} \\
            -\frac{4E_2 a^2}{h} ,~~~~     & h>\frac{\sqrt{2}}{2}
           \end{array} \right .
\label{eq:m1-2}
\end{equation} 
Here $|0\rangle$ represents the ground state, which is $|\Psi_1\rangle$ for $h<\frac{\sqrt{2}}{2}$ and $|\Psi_2\rangle$ for $h>\frac{\sqrt{2}}{2}$. The result is illustrated in Fig. 4(c). At $h=\frac{\sqrt{2}}{2}-0^+$, the magnetizaiton $m_1$ is zero, and at $h=\frac{\sqrt{2}}{2}+0^+$,  $m_1=\frac{2}{3}\sqrt{2}$. Hence, the surface magnetization $m_1$  exhibits a discontinuity at the transition point and undergoes a finite jump. The susceptibility  $\frac{\partial m_1}{\partial h}$ becomes infinite at the phase transition point:
\begin{equation}
\frac{\partial m_1}{\partial h}|_{h=\sqrt{2}/2}=\infty.
\label{eq:suscep}
\end{equation}

From Equation (\ref{eq:m1m2}), we obtain $m_1=-m_2$, where $m_2$ represents the magnetization of spin 2. Consequently, the susceptibility $\frac{\partial m_2}{\partial h}$ also becomes infinite at the phase transition point. Numerical calculations for larger lattices reveal the emergence of an anti-symmetric magnetization profile with respect to the center of the lattice.

The non-analyticity is also observed in the entanglement entropy, which captures the presence of non-classical correlations between distinct systems \cite{kitaev, yuste}. If we consider the two-site model as two separate systems, with site 1 denoting system A and site 2 denoting system B, we define the reduced density matrix as follows:
\begin{equation}
\rho_A=Tr_B|0\rangle\langle 0|
\end{equation}
Here, $|0\rangle$ represents the ground state under consideration. The entanglement entropy is then defined as:
\begin{equation}
S_E(A)=-Tr(\rho_A \log_2 \rho_A)
\end{equation}
For $h_L=h_R=h$, the ground state is described by {\small $|\Psi_1\rangle$} in Equation (\ref{eq:psi12}) for {\small $h<\frac{\sqrt{2}}{2}$}, and by {\small $|\Psi_2\rangle$} in Equation (\ref{eq:psi12}) for $h>\frac{\sqrt{2}}{2}$. After performing the necessary calculations, the results are as follows:
\begin{equation}
S_E(A)=\left \{
\begin{array}{ll}
 1, & h=\frac{\sqrt{2}}{2}-0^+ \\
 \log_2 6- \frac{4\sqrt{2}}{3} \log_2 (1+\sqrt{2}), & h=\frac{\sqrt{2}}{2}+0^+ .
 \end{array}
 \right.
\label{eq:entropy-ab}
\end{equation}
The value {\small $\log_2 6- \frac{4\sqrt{2}}{3} \log_2 (1+\sqrt{2})=0.1872\cdots$} reflects the non-analytic behavior of the entanglement entropy at the transition point.

The above results obtained from the two-site model serve as a benchmark for subsequent analysis and numerical investigations. Moreover, in this simple model, we illustrate the common properties observed for even values of $N$: the discontinuity in $\frac{\partial \Delta}{\partial h}$ as indicated in Equation (\ref{eq:delta-ab}), and the infinite value of $\frac{\partial m_i}{\partial h}, i=1,2$ as expressed in Equation (\ref{eq:suscep}).

\begin{figure}i
\includegraphics[width=0.5\textwidth]{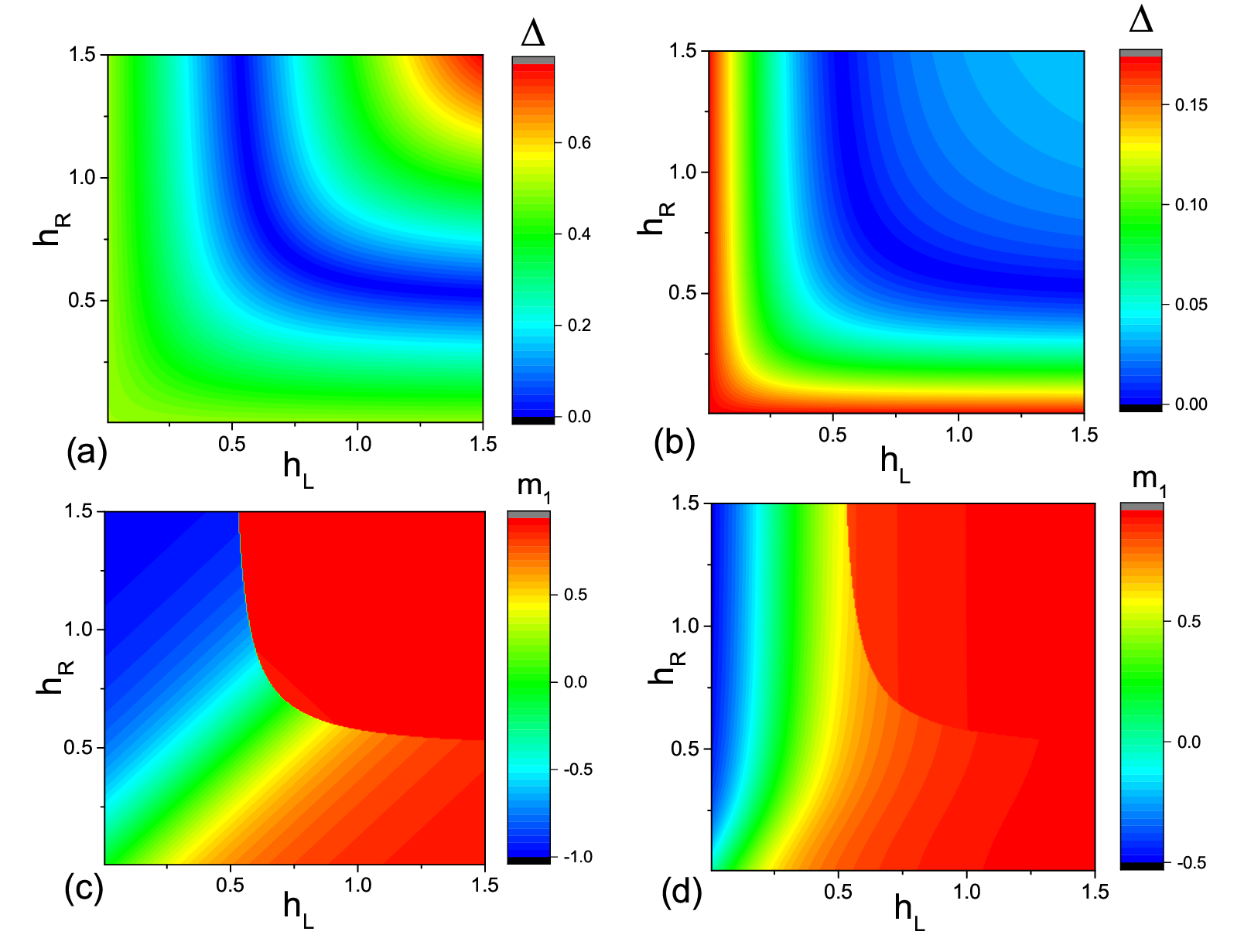}
\caption{(a) Color map representing the energy gap for $N=2$. (b) Color map representing the energy gap for $N=8$. (a) Color map illustrating the surface magnetization $m_1$ for $N=2$. Color map illustrating the surface magnetization $m_1$ for $N=8$}
\end{figure}

\subsection{Numerical Results for Lattices with Even Sites ($N>2$)}

For lattice models with an even number of sites $N=2n$ and $n>1$, effective solutions can be obtained using the Jordan-Wigner transformation (see Appendix A). Figure 2 illustrates the color maps of the energy gap and surface magnetization $m_i$ for $N=2$ and $N=8$ under different surface fields $h_L$ and $h_R$. From Figures 2(a) and 2(b), it is evident that the energy gaps close along the curve described by Equation (\ref{eq:pt}). Furthermore, Figures 2(c) and 2(d) indicate that the surface magnetization $m_1$ undergoes an abrupt change at the phase boundary defined by Equation (\ref{eq:pt}). A rigorous proof of this phase boundary for lattices with all even sites is presented in Appendix B. Thus, Figure 1(d) depicts the phase diagram, where the red solid curve follows Equation (\ref{eq:pt}).

The energy gap $\Delta$ is shown for $N=4$,$N=8$ and $N=40$ in Figures 3(a), 3(b), and 3(c), respectively. These plots demonstrate that the phase transition occurs at the point $h_L=h_R=\frac{\sqrt{2}}{2}$, where the energy gap closes. Close to the phase transition point, the energy gap curve exhibits a sharp cusp, indicating singularity with a discontinuous derivative $\frac{\partial \Delta}{\partial h}$, similar to the two-site model shown in Equation (\ref{eq:delta-ab}).

\begin{figure}
\includegraphics[width=0.5\textwidth]{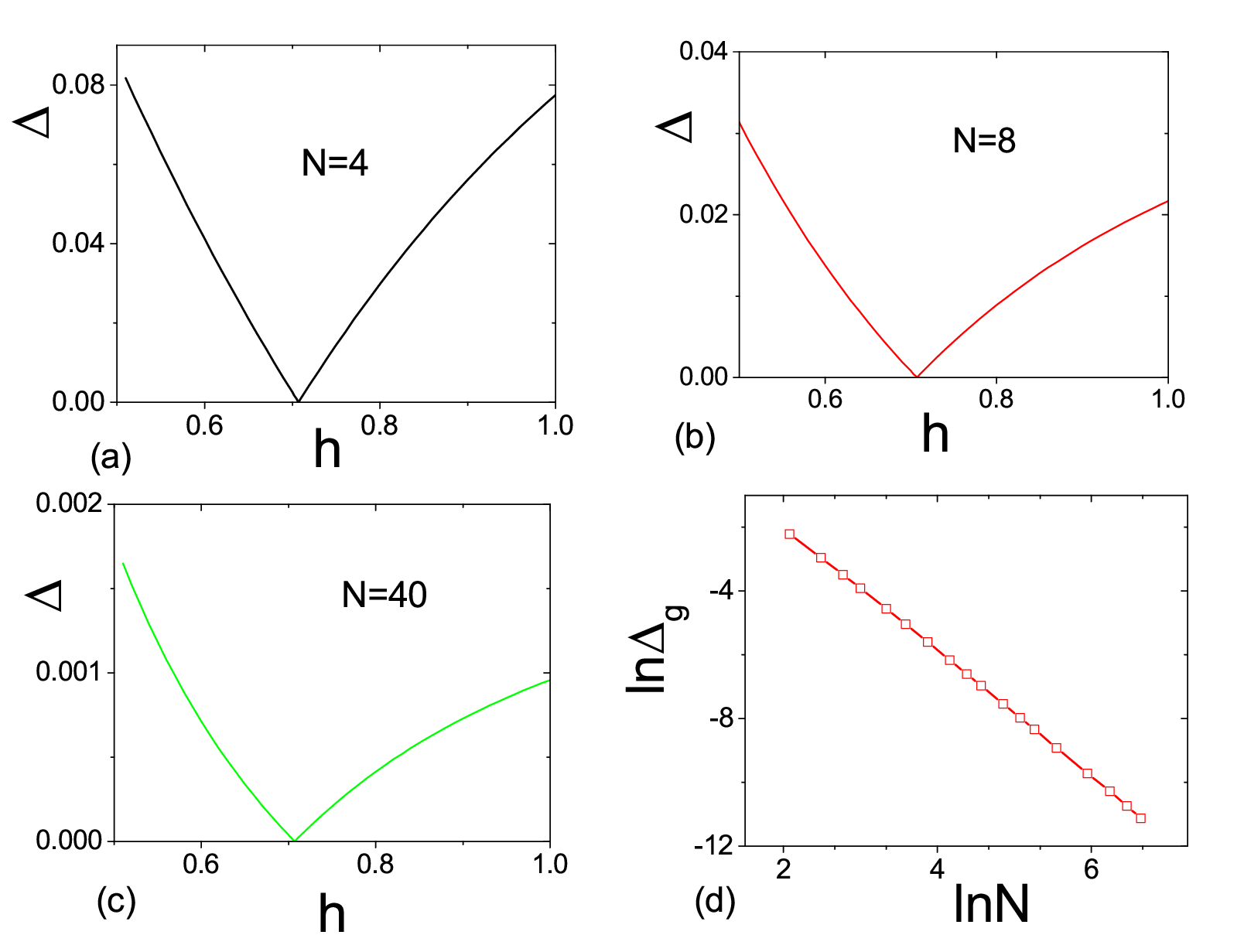}
\caption{(a), (b), and (c) display the energy gap $\Delta$ for $N=4$, $N=8$ and $N=40$, respectively. The energy gap exhibits a sharp cusp at 
$h_L=h_R=\frac{\sqrt{2}}{2}$ . (d) Illustrates the jump of $\frac{\partial \Delta}{\partial h}$  at the transition point as a function of lattice size $N$. Here, $\Delta_g$is defined by Equation (\ref{eq:delta-g}).}
\end{figure}

We observe a discontinuity in $\frac{\partial \Delta}{\partial h}$  at the transition point. The jump is defined as
\begin{equation}
\Delta_g = \left.\frac{\partial \Delta}{\partial h}\right|_{h=h_c+0^+} - \left.\frac{\partial \Delta}{\partial h}\right|_{h=h_c-0^+},
\label{eq:delta-g}
\end{equation}
where $h_c=\frac{\sqrt{2}}{2}$ in this context. The decrease in the jump $\Delta_g$ with increasing lattice size $N$ is evident from Figures 3(a), 3(b), and 3(c). We calculated this jump for $10<N<800$, and the results are presented in Figure 3(d). By fitting the data in Figure 3(d), we find that
\begin{equation}
\Delta_g \propto N^{-1.959(6)}.
\end{equation}
This suggests that the singularity in the energy gap diminishes as the lattice size increases, eventually vanishing in the thermodynamic limit.

\begin{figure}
\includegraphics[width=0.5\textwidth]{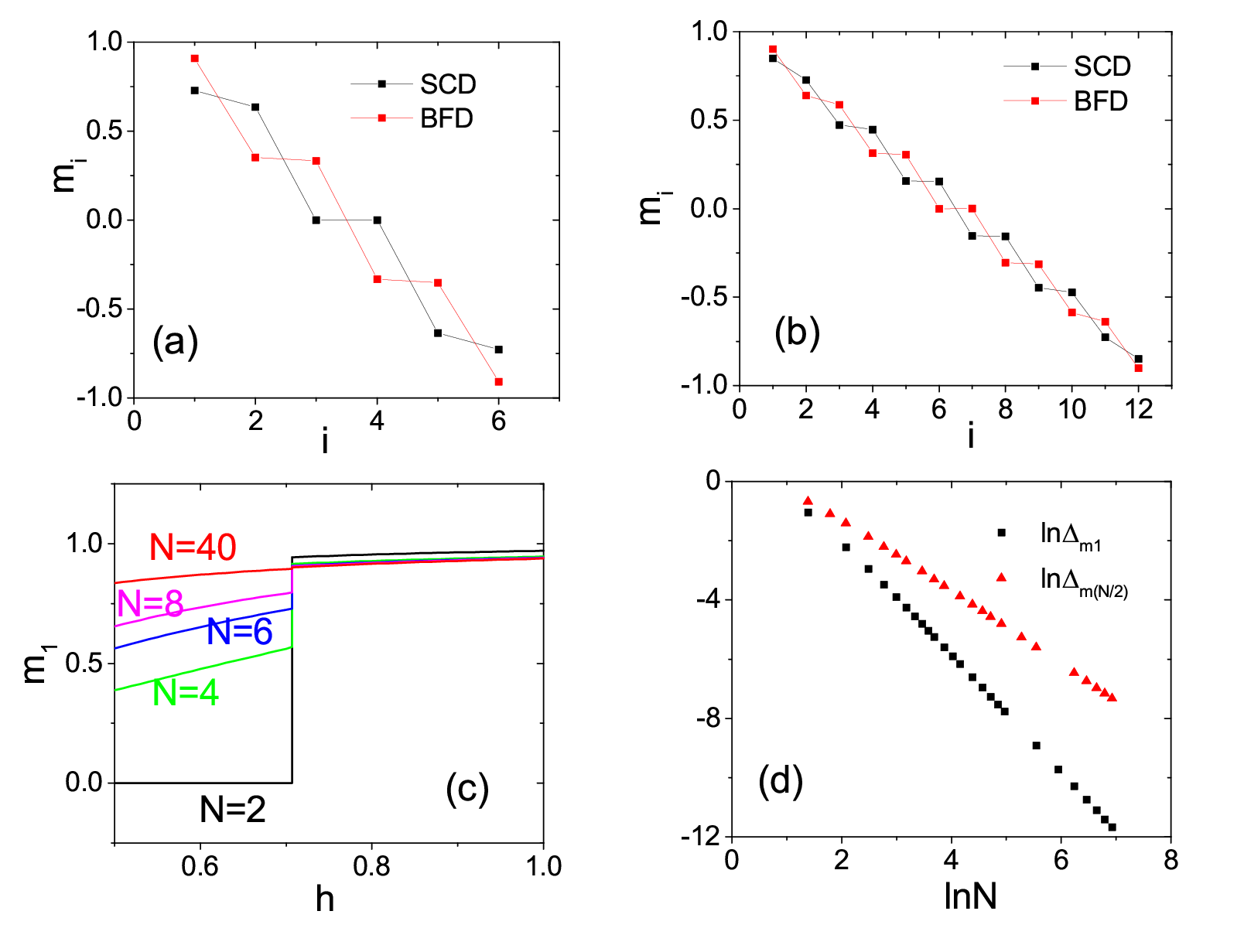}
\caption{(a)Magnetization profiles for $N=6$ just above and below the transition point $h_L=h_R=\frac{\sqrt{2}}{2}$. (b)Magnetization profiles for $N=12$ just above and below $h_L=h_R=\frac{\sqrt{2}}{2}$. (c) Surface magnetization $m_1$ as a function of $h$. (d) Jump amplitudes of $m_1$ and $m_{\frac{N}{2}}$ at the transition for different $N$. }

\end{figure}

The first-order phase transition is directly observable in the magnetization along the X-direction at each site, which is defined as follows:

\begin{equation}
m_i\equiv \langle 0|\sigma_i^x|0\rangle,
\end{equation}

where the Jordan-Wigner transformation can be utilized to calculate the magnetization $m_i$ \cite{wu2}. Fig.4(a) and Fig.4(b) showcase the magnetization profiles near the transition point for $N=6$ and $N=12$ respectively. The SCD phase is represented by the black lines and squares, while the BFD phase is denoted by the red lines and squares. Here, the SCD phase occurs at $h_L=h_R=\frac{\sqrt{2}}{2}-10^{-12}$ in double precision, and the BFD phase occurs at $h_L=h_R=\frac{\sqrt{2}}{2}+10^{-12}$. These values are slightly above and below the transition point $h_L=h_R=\frac{\sqrt{2}}{2}$. As we can observe, the magnetization at each site differs above and below the transition point.

As an example, we demonstrate the variations in surface magnetization along the line $h_L=h_R=h$ for different sizes, such as $N=2,4,6,8,40$, in Fig.4(c). Prominent jumps occur at $h=1/\sqrt{2}=0.707\cdots$ with decreasing jump amplitudes as the size $N$ increases. At the transition point, every site's magnetization experiences a jump. We quantify the jump amplitude for the $i$th spin as follows:

\begin{equation}
\Delta_{mi}=|(m_i|_{h=\frac{1}{\sqrt{2}}+0^+}-m_i|_{h=\frac{1}{\sqrt{2}}-0^+})|,
\label{eq:jump}
\end{equation}

According to the aforementioned discussion, the jump amplitude of the surface magnetization for $N=2$ can be read from Fig.4(c) as $\Delta_{m1}=2\sqrt{2}/3$.

We have computed the surface magnetization jump $\Delta_{m1}$ and the magnetization jump of the middle spin ($\frac{N}{2}$th spin) $\Delta_{m\frac{N}{2}}$ for various sizes, as shown in Fig.4(d). By fitting the obtained numerical results in Fig.4(d), we find that:

\begin{equation}
\Delta_{m1}\sim N^{-1.986(1)},~~~\Delta_{m\frac{N}{2}}\sim N^{-1.237(2)},
\label{eq:decrease}
\end{equation}

for $N\gg 1$ and $N$ being even.

It is worth noting that there is no finite-size rounding in the phase transition. At the transition point, the energy gap becomes zero, and the magnetization at each site undergoes a discontinuous change along the phase boundary without any finite-size rounding. However, as the lattice size approaches infinity, the jumps in magnetizations vanish.

In the exact solution for $N=2$, we reveal the parity in the two phases through Eq. (\ref{eq:parity1}). Now, we argue that for $N>2$ and $N$ being even, the parity remains the same as $N=2$ in the two phases. Let's consider the point where $h_L=h_R=0$, which corresponds to the original point in the phase diagram in Fig.1(d). At this point, the ground state must possess even parity. If we remove the term $\sum_i \sigma_i^y \sigma_{i+1}^y$ from the Hamiltonian, there exist two degenerate ground states: $|1,1,\cdots,1\rangle$ (all $s_i$ being positive) and $-1,-1,\cdots,-1\rangle$ (all $s_i$ being negative). Both of these states have even parity since $N$ is even. Now, let's imagine gradually introducing the term $\sum_i \sigma_i^y \sigma_{i+1}^y$ back into the Hamiltonian. In this process, the parity of the ground state should not change. 

\section{Lattices with an Odd Number of Sites}

\subsection{Exact Solution for $N=3$}

We have obtained the exact solution for the three-site model. Similarly, we employ a basis in which $\sigma_i^x$ is diagonal. With three spins, there exist eight states. The four states $|\psi_1\rangle =|1,1,1\rangle$, $|\psi_2\rangle=|1,-1,-1\rangle$, $|\psi_3\rangle =|-1,1,-1\rangle$, and $|\psi_4\rangle=|-1,-1,1\rangle$ exhibit even parity, while the remaining four states $|\psi_5\rangle =|1,1,-1\rangle$, $|\psi_6\rangle=|1,-1,1\rangle$, $|\psi_7\rangle =|-1,1,1\rangle$, and $|\psi_8\rangle=|-1,-1,-1\rangle$ possess odd parity. In this specific basis, the Hamiltonian can be expressed as:

\begin{equation}
H=\left ( \begin{array}{cc}
            H^{(+)} &  0 \\
            0      &H^{(-)}
           \end{array} \right ),
\end{equation}

where $H^{(+)}$ represents the matrix corresponding to the subspace of even parity:

\begin{equation}
H^{(+)}=-\frac{1}{4}\left ( \begin{array}{cccc}
            a_+ &  1 & 0 & 1 \\
            1 &  b & 1 & 0 \\
            0 &  1 & -a_+ & 1 \\
            1 &  0 & 1 & -b 
           \end{array} \right ),
\end{equation}

with $a_+=2(1+h_L-h_R)$ and $b=2(h_L+h_R)$. Similarly, $H^{(-)}$ represents the matrix corresponding to the subspace of odd parity:

\begin{equation}
H^{(-)}=-\frac{1}{4}\left ( \begin{array}{cccc}
            b &  1 & 0 & 1 \\
            1 &  -a_- & 1 & 0 \\
            0 &  1 & -b & 1 \\
            1 &  0 & 1 & a_-
           \end{array} \right ),
\end{equation}

with $a_-=2(1+h_R-h_L)$. The eigenvalues of the Hamiltonian in the even parity subspace, $E^{(+)}$, are given by:

\begin{equation}
E^{(+)}=\pm \frac{1}{4\sqrt{2}}\sqrt{A_+\pm\sqrt{A_+^2-4a_+^2b^2}},
\end{equation}

where $A_+=a^2+b^2+4$. Similarly, the eigenvalues of the Hamiltonian in the odd parity subspace, $E^{(-)}$, are given by:

\begin{equation}
E^{(-)}=\pm \frac{1}{4\sqrt{2}}\sqrt{A_-\pm\sqrt{A_-^2-4a_-^2b^2}},
\end{equation}

where $A_-=a_-^2+b^2+4$.

By comparing the aforementioned solutions, we can determine that when $h_L>h_R$, the ground state corresponds to the minimum of $E^{(+)}$ and it possesses even parity, whereas when $h_L>h_R$, the ground state corresponds to the minimum of $E^{(-)}$ and it possesses odd parity. The ground state energy can be expressed as:

\begin{equation}
E_0=\left \{ \begin{array}{cc}
                 -\frac{1}{4\sqrt{2}}\sqrt{A_+\sqrt{A_+^2+4a_+^2b^2}}, & h_L>h_R \\
                 -\frac{1}{4\sqrt{2}}\sqrt{A_-\sqrt{A_-^2+4a_-^2b^2}}, & h_L<h_R
                \end{array} \right.
\end{equation}

Similarly, the parity of the phase is given by:

\begin{equation}
P_x=\left \{ \begin{array}{cc}
                 1,~~~ & h_L>h_R \\
                 -1, ~~~ & h_L<h_R
                \end{array} \right.
\label{eq:parity2}
\end{equation}

The phase boundary is defined by:

\begin{equation}
h_L=h_R,
\label{eq:pt1}
\end{equation}

which corresponds to the diagonal line depicted in Figure 1(c). At this point, the energy gap exhibits a sharp cusp, as demonstrated in Figure 6(a).

\subsection{Numerical Results for Lattices with Odd Sites $N>3$}

\begin{figure}
\includegraphics[width=0.5\textwidth]{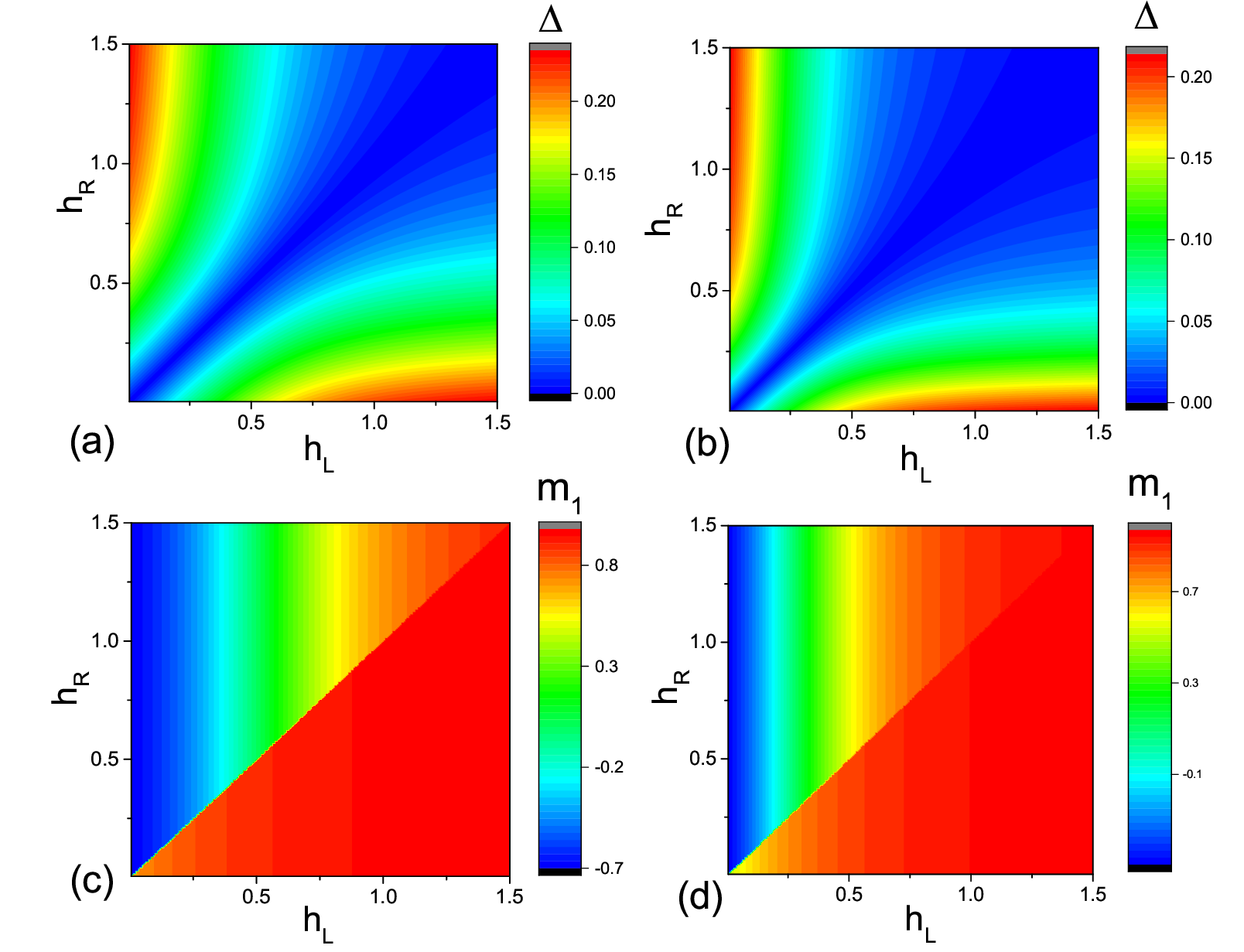}
\caption{(a) Color map illustrating the energy gap for $N=3$. (b) Color map illustrating the energy gap for $N=7$. (a) Color map illustrating the surface magnetization $m_1$ for $N=3$. Color map illustrating the surface magnetization $m_1$ for $N=7$.}

\end{figure}

Utilizing the Jordan-Wigner transformation \cite{wu2}, we have computed the energy gap and magnetization profile. It has been rigorously proven in Appendix C that the phase boundary for lattices with odd sites $N>3$ remains the same. Figure 5 presents the color map of the energy gap and the surface magnetization $m_1$ for $N=3,7$ under various surface fields $h_L,h_R$. Observing Fig. 5(a) and 5(b), we can infer that the energy gaps close following the curve described by Eq. (\ref{eq:pt1}). Fig. 5(c) and 5(d) demonstrate that the surface magnetization $m_1$ undergoes an abrupt change precisely at the phase boundary indicated by Eq. (\ref{eq:pt1}). Consequently, we obtain the phase diagram presented in Fig. 1(c), wherein the red solid curve corresponds to Eq. (\ref{eq:pt1}).

\begin{figure}
\includegraphics[width=0.5\textwidth]{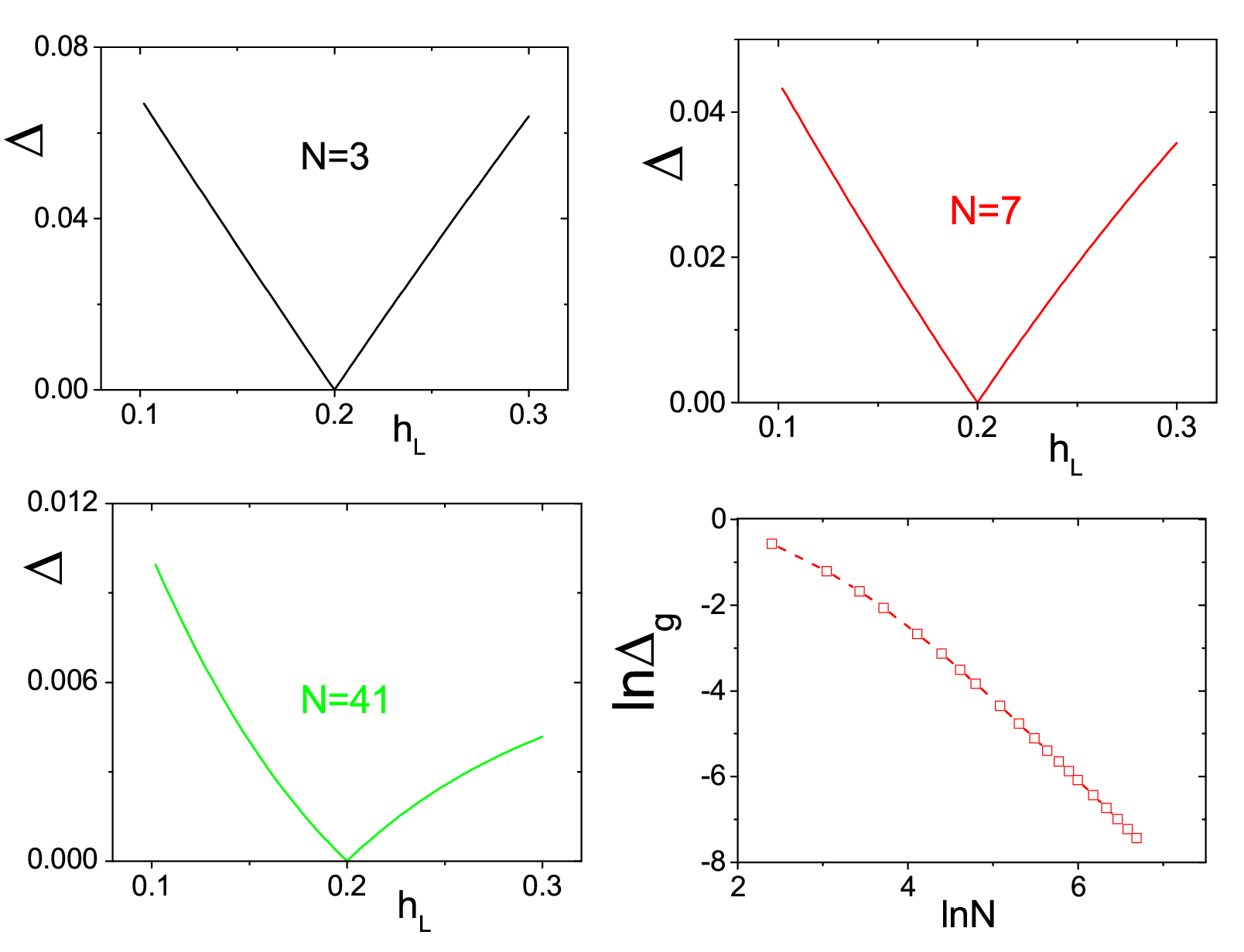}
\caption{(a), (b), (c) Energy gap $\Delta$ for $N=3,7,41$, respectively. (d) Jump in $\frac{\partial \Delta}{\partial h_R}$ at $h_R=0.2=h_L$ for different sizes. $\Delta_g$ is defined by Eq. (\ref{eq:delta-g2}).}

\end{figure}

Figures 6(a), 6(b), and 6(c) display the energy gap $\Delta$ for $N=3,7,41$, respectively. For these cases, we set $h_R=0.2$ and vary $h_L$ from $0.1$ to $0.3$. The phase transition point occurs when $h_R=h_L=0.2$, where the energy gap vanishes. In the vicinity of the phase transition point, the energy gap curve exhibits a sharp cusp, indicating singularity and discontinuity in $\frac{\partial \Delta}{\partial h_L}$, similar to the two-site model presented in Eq. (\ref{eq:delta-ab}).

In a similar fashion, we define a jump as
\begin{equation}
\Delta_g=\frac{\partial \Delta}{\partial h_L}|_{h_R=h_c+0^+}-\frac{\partial \Delta}{\partial h_L}|_{h_R=h_c-0^+},
\label{eq:delta-g2}
\end{equation}
where $h_c=h_R=0.2$ in this case. From Figs. 6(a), 6(b), and 6(c), it can be observed that the jump $\Delta_g$ decreases with increasing lattice size. We have calculated this jump for $10<N<800$, and the results are presented in Fig. 6(d). By fitting the data in Fig. 6(d), we find that
\begin{equation}
\Delta_g \propto N^{-1.952(3)},
\end{equation}
for large odd values of $N$. This indicates that the singularity in the energy gap diminishes as the lattice size increases. In the thermodynamic limit, the singularity in the energy gap should ultimately vanish.

\begin{figure}
\includegraphics[width=0.5\textwidth]{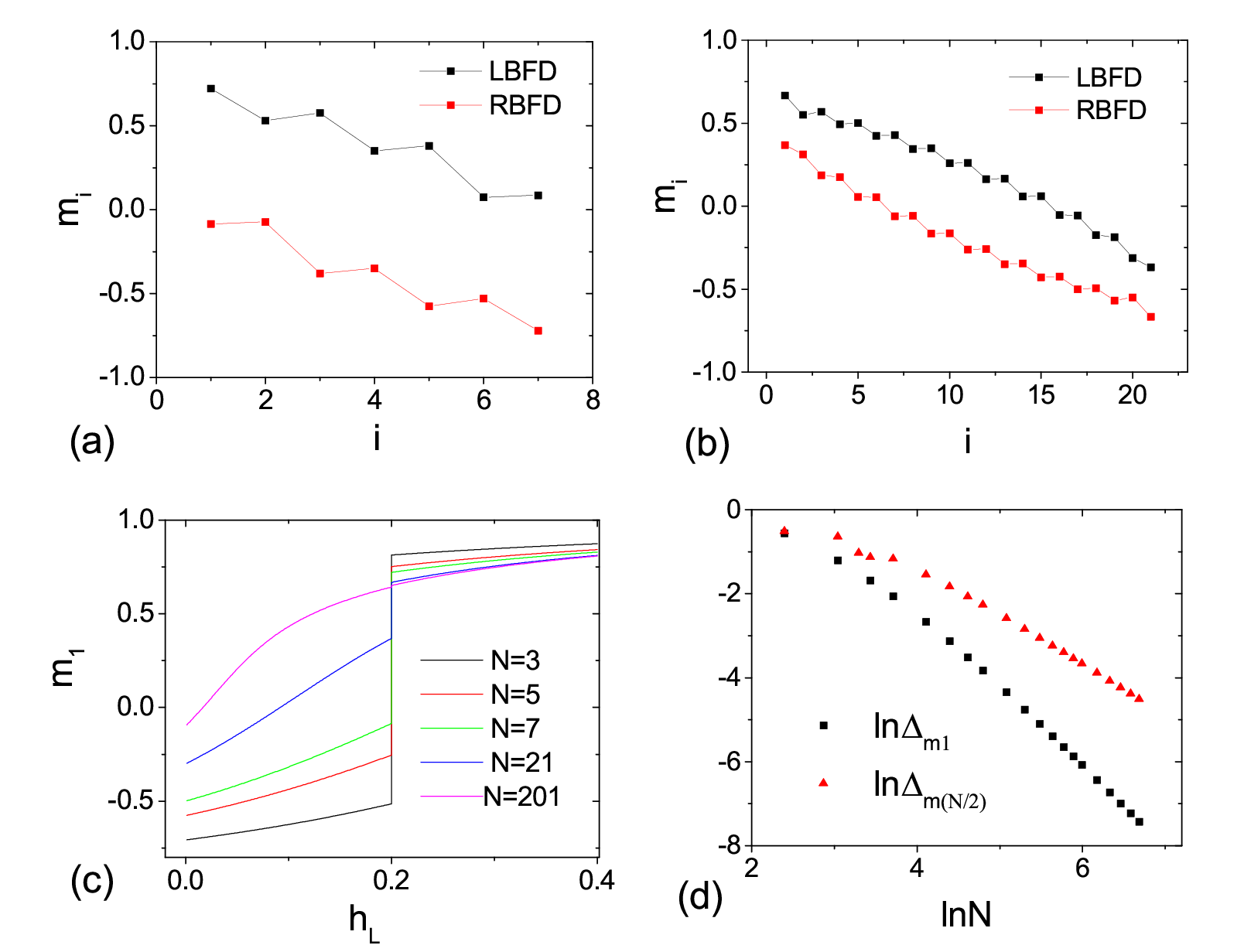}
\caption{(a) Magnetization profiles for $N=7$ just above and below the transition point $h_L=0.2$ with $h_R=0.2$. (b) Magnetization profiles for $N=21$ just above and below $h_L=0.2$ with $h_R=0.2$. (c) Surface magnetization $m_1$ versus $h_L$ with $h_R=0.2$. (d) Jump amplitudes of $m_1$ and $m_{\frac{N+1}{2}}$ at the transition for different $N$.}

\end{figure}

Magnetization $m_i$ can be computed using the Jordan-Wigner transformation \cite{wu2}. We present the magnetization profiles for $N=7,21$ in the vicinity of the transition point with $h_R=0.2$ in Fig. 7(a) and Fig. 7(b), respectively. The LBFD phase is indicated by the black line and square, while the RBFD phase is indicated by the red line and square. The LBFD phase is obtained at $h_L=0.2-10^{-12}$ with double precision, and the RBFD phase is obtained at $h_L=0.2+10^{-12}$. These values are slightly above and below the transition point $h_R=h_L=0.2$. As shown, the magnetization at each site exhibits distinct behavior below and above the transition point.

In a similar fashion, we define a jump in magnetizaiton as
\begin{equation}
\Delta_mi=m_i |_{h_R=h_c+0^+}-m_i |_{h_R=h_c-0^+},
\label{eq:delta-m2}
\end{equation}
where $h_c=h_L=0.2$ in this case.

Figure 7(c) illustrates the surface magnetization for $n=3,5,7,21,201$, with $h_L=0.2$ fixed, thereby setting the phase transition point at $h_R=0.2$. The surface magnetization $m_1$ experiences a jump at the transition point.

As representatives of the phenomenon, we compute the jump amplitudes at the first site and the $(N+1)/2$th site, as shown in Fig. 7(d). By fitting the numerical results in Fig. 7(d), we find the following scaling behavior:
\begin{equation}
\Delta_{m1}\sim N^{-1.953(2)},~~~\Delta_{m\frac{N+1}{2}}\sim N^{-1.217(2)},
\label{eq:decrease1}
\end{equation}
for large odd values of $N$. These results indicate that the magnetization jumps decrease with increasing lattice size according to Eq. (\ref{eq:decrease}). When considering these jumps as characteristic of a first-order phase transition, they become weaker as the lattice size grows. In the limit as $N\rightarrow \infty$, these jumps should vanish. This behavior is contrary to the conventional notion of phase transitions, which typically occur only in the true sense as $N\rightarrow \infty$.

In the exact solution for $N=3$, we demonstrate the parity in the two phases in Eq. (\ref{eq:parity2}). Here, we argue that for $N>3$ and $N$ being odd, the parity remains the same as for $N=3$ in the two phases. Consider the point $h_L\gg 1, h_R=0$, corresponding to the points on the horizontal axis in the phase diagram shown in Fig. 1(c). At these points, the ground state must exhibit even parity. In the absence of the term $\sum_i \sigma_i^y \sigma_{i+1}^y$ in the Hamiltonian, the ground state should be $|1,1,\cdots,1\rangle$ (with all $s_i$ being positive), which possesses even parity. If we gradually introduce the term $\sum_i \sigma_i^y \sigma_{i+1}^y$, the parity of the ground state should remain unchanged. Therefore, for $h_L\gg 1,h_R=0$, the ground state retains even parity. Since a change in symmetry implies the occurrence of a phase transition, we conclude that for $N>3$ and $N$ being odd, the parity remains the same as for $N=3$ in the two phases.

\section{Numerical Method}

In this study, we employ well-established theories \cite{bariev,hinrichsen,bilstein,vicari,wu1,wu2} to transform the diagonalization problem into an effective Hamiltonian by introducing an additional spin on both the left and right sides. The resulting Hamiltonian is given by:

\begin{equation}
H_e  = -\frac{1}{4}\sum_{i=1}^{N-1}(\sigma_i^x \sigma_{i+1}^x+\sigma_i^y \sigma_{i+1}^y) -\frac{1}{2}h_L \sigma^x_0 \sigma^x_1-\frac{1}{2}h_R \sigma^x_N\sigma^x_{N+1}. 
\label{eq:hamiltonian}
\end{equation}

In this formulation, the spins $\sigma_0^x$ and $\sigma_{N+1}^x$ commute with the Hamiltonian, allowing them to be diagonalized simultaneously. The Hilbert space can be divided into four sectors labeled as $(1,1),(1,-1),(-1,1),(-1,-1)$, where $(s_0,s_{N+1})$ represent the eigenvalues of $\sigma^x_0$ and $\sigma^x_{N+1}$. By restricting $H_e$ to these four sectors, we obtain the Hamiltonian $H$ representing different cases characterized by the signs of $h_L$ and $h_R$ \cite{vicari}. Notably, the first-order phase transition occurs in the $(1,-1)$ sector, which is the focus of our investigation.

To solve the effective Hamiltonian $H_e$, we utilize the Jordan-Wigner transformation \cite{lieb,wu2}. In Appendix C and D, we provide a proof of the phase boundaries Eq. (\ref{eq:pt}) and (\ref{eq:pt1}) based on the aforementioned effective Hamiltonian.

Furthermore, it is worth noting that the Hamiltonian can be decomposed into two decoupled subsystems, which is evident from the Jordan-Wigner transformation shown in Appendices C and D.

{\sl Dynamics} - As we delve into the dynamics of phase transitions, we encounter an important subject \cite{kibble,zurek,zurek1,damski}. In our model, governed by the properties described in Eq. (\ref{eq:oddeven}) and Eq. (\ref{eq:commute}), the Hilbert space is intricately divided into two distinct subspaces. Consequently, modifying the boundary fields $h_L$ and $h_R$ might prevent the system from reaching its ground state, impeding the realization of a phase transition even by abruptly adjusting the boundary fields to the transition point.

To illustrate this point, let's consider the two-site model with $h_L=h_R=h$. In this scenario, the ground state is given by $|\Psi_1\rangle$ in Eq. (\ref{eq:psi12}) for $h<\frac{\sqrt{2}}{2}$, and $|\Psi_2\rangle$ in Eq. (\ref{eq:psi12}) for $h>\frac{\sqrt{2}}{2}$. Suppose the initial boundary fields $h$ are small, and the system is in the ground state $|\Psi_1\rangle$. If we gradually increase $h$ adiabatically to go beyond the transition point at $h=\frac{\sqrt{2}}{2}$, accessing the actual ground state $|\Psi_2\rangle$ becomes unattainable. Since the parities of $|\Psi_1\rangle$ and $|\Psi_2\rangle$ differ, and parity is conserved, their hybridization is hindered.

To realize the phase transition, we need to introduce additional terms in the Hamiltonian Eq. (\ref{eq:oddeven}) to couple the spins on odd and even sites. In the two-site model, we incorporate a term $g\sigma_1^z$ into the Hamiltonian, with $g$ being a small value, to facilitate the hybridization between $|\Psi_1\rangle$ and $|\Psi_2\rangle$ as mentioned above. Afterwards, the boundary fields can be quenched to values above the phase transition point, followed by tuning $g$ to zero. This approach enables the realization of the phase transition.

\section{Summary and Acknowledgments}

In this study, we have explored the behavior of the XY model with opposite boundary fields directed in the X direction. Remarkably, we have discovered a first-order few-body phase transition in this system. Notably, the phase diagrams differ for lattices with even and odd numbers of sites, which adds to the intriguing nature of this model. These unique properties, including the qualitative differences between even and odd site numbers and the existence of two distinct Hilbert subspaces, hint at potential applications in the realms of quantum information and quantum computation.

We would like to express our gratitude to Wenan Guo and the SGI in the Department of Physics at Beijing Normal University for providing us with valuable computing resources.

\appendix

\section{The Jordan-Wigner Transformation}

To investigate the spectrum of the Hamiltonian in Eq. (\ref{eq:hamiltonian}), we employ the Jordan-Wigner transformation and introduce fermionic operators \cite{lieb}:

\begin{eqnarray}
\sigma_i^{+} & = & (-1)^{i-1}\prod_{j=0}^{i-1}(2c^{\dagger}_jc_j-1)c^{\dagger}_i, \nonumber \\
\sigma_i^{-} & = & (-1)^{i-1}\prod_{j=0}^{i-1}(2c^{\dagger}_jc_j-1)c_i.
\end{eqnarray}

Here, $\sigma^{\pm}=(\sigma^x \pm \mathfrak{i} \sigma^y)/2$ ($\mathfrak{i}$ denotes the imaginary unit). The Hamiltonian can be expressed as:

\begin{equation}
H_e=-gN+\sum_{i,j=0}^{N+1}(c_i^{\dagger}{\bf A}_{ij}c_j+\frac{1}{2}c_i^{\dagger}{\bf B}_{ij}c_{j}^{\dagger}-\frac{1}{2}c_i {\bf B}_{ij}c_j),
\end{equation}

where ${\bf A}$ and ${\bf B}$ are symmetric and anti-symmetric matrices, respectively. To elucidate the formulation, we explicitly present the matrix elements for $N=3$:

\begin{equation}
{\bf A}=\left ( \begin{array}{ccccc}
            0      &  - h_L &  0 &  0 &  0 \\
            -h_L   &  0     & -1 &  0 &  0  \\
            0      &  -1    &  0 & -1 &  0  \\
            0      &   0    & -1 &  0 & -h_R \\
            0      &   0    &  0 & -h_R&  0
           \end{array} \right )
\end{equation}

\begin{equation}
{\bf B}=\left ( \begin{array}{ccccc}
            0      &  -h_L &  0 &  0 &  0 \\
            h_L    &  0    &  0 &  0 &  0  \\
            0      &  0    &  0 &  0 &  0  \\
            0      &  0    &  0 &  0 & -h_R \\
            0      &  0    &  0 &  h_R &  0
           \end{array} \right ).
\end{equation}

By introducing new canonical fermionic variables \cite{lieb}, the Hamiltonian can be diagonalized using a Bogoliubov transformation:

\begin{equation}
\eta_k=g_{k,i}c_i+h_{k,i}c_i^{\dagger}.
\end{equation}

The coefficients $g_{k,i}$ and $h_{k,i}$ satisfy the following equations:

\begin{equation}
g_{ki}=\frac{\phi_{k,i}+\psi_{k,i}}{2}, \hskip 0.5cm h_{ki}=\frac{\phi_{k,i}-\psi_{k,i}}{2},
\end{equation}

where $\psi_k$ is the eigenvector of the matrix 

\begin{equation}
{\bf C}\equiv ({\bf A}+{\bf B})({\bf A}-{\bf B}),~~~~{\bf C}\psi_k=\varepsilon_k^2 \psi_k,
\label{eq:AB}
\end{equation}

and 

\begin{equation}
\phi_k=({\bf A}-{\bf B})\psi_k /\varepsilon_k.
\label{eq:fk}
\end{equation}

In the above equations, $\varepsilon_k \neq 0$. Notably, ${\bf C}$ possesses a zero eigenvalue $\varepsilon_0=0$, addressed in reference \cite{vicari}. For the convenience of our readers, we present the explicit matrix elements of ${\bf C}$ for $N=3$:

\begin{equation}
{\bf C}=\frac{1}{4}\left ( \begin{array}{ccccc}
            4h_L^2  &  0  &  2h_L &  0 &  0 \\
            0       &  1   &  0 &  1 &  0  \\
            2h_L    &  0   &  2 &  0 &  0  \\
            0       &  1   &  0 &  1+4h_R^2 & 0 \\
            0       &  0   &  0 &  0 &  0
           \end{array} \right ).
\end{equation}

One can observe that all the elements in the $N+2$th column and $N+2$th row of $C$ are zero. Thus, ${\bf C}$ possesses a zero eigenvalue $\varepsilon_0=0$. Due to this zero mode, the spectrum exhibits a twofold degeneracy, arising from the $Z_2$ global symmetry of Hamiltonian $H_e$. However, this zero mode is unrelated to the spectrum of the Hamiltonian $H$. Only the non-zero modes are pertinent. There exist $N+1$ non-zero modes labeled as $k=1,2,\cdots,N+1$. Notably, $0=\varepsilon_0<\varepsilon_1<\varepsilon_2 \cdots$.

It is worth noting that in the subsequent discussions, we exclude the ${(N+2)}$th column and ${(N+2)}$th row from the matrix $C$ since their elements are zero.

\section{The Phase Boundary for Lattices with Even Sites}

The matrix $C$ for a general $XY$ model has been extensively discussed in \cite{wu2}. To provide clarity, we present the matrix $C$ for models with two, four, and six sites. 

For the two-site model, Matrix $C$ is defined as follows:
\begin{equation}
C_3=\frac{1}{4}\left ( \begin{array}{ccc}
           4h_L^2   & 0   &  2h_L \\
           0      & 1   &   0 \\
           2h_L     & 0   &  1+4h_R^2
         \end{array} \right ).
\end{equation}

For the four-site model, Matrix $C$ is given by:
\begin{equation}
C_5=\frac{1}{4}\left ( \begin{array}{ccccc}
           4h_L^2   & 0   &  2h_L  & 0  & 0\\
           0      & 1   &   0  & 1  & 0\\
           2h_L     & 0   &   2  & 0  & 1\\
           0      & 1   &   0  & 2  & 0\\
           0      & 0   &   1  & 0  &  1+4h_R^2
         \end{array} \right ).
\end{equation}

And for the six-site model, Matrix $C$ is expressed as follows:
\begin{equation}
C_7=\frac{1}{4}\left ( \begin{array}{ccccccc}
           4h_L^2   & 0   &  2h_L  & 0   & 0    & 0  & 0\\
           0        & 1   &   0    & 1   & 0    & 0  & 0\\
           2h_L     & 0   &   2    & 0   & 1    & 0  & 0\\
            0       & 1   & 0      & 2   &   0  & 1  & 0 \\
            0       & 0   & 1      & 0   &   2  & 0  & 1 \\
            0       & 0   & 0      & 1   &   0  & 2  & 0 \\
           0        & 0   & 0      & 0   & 1   & 0   & 1+4h_R^2
         \end{array} \right ).
\end{equation}

In the above expressions, we omit the $(N+2)$th row and column. Similar representations can be derived for matrices with more sites.

It is worth noting that the matrix elements satisfy $C_{2j,2k+1}=C_{2k+1,2j}=0$, where $j$ and $k$ are non-negative integers. This implies that odd sites only couple with odd sites, and even sites only couple with even sites. Consequently, the Hamiltonian can be divided into two parts: one associated with spins on odd sites and the other with spins on even sites. This property is demonstrated through a duality transformation in Eq. (9) of the referenced paper.

We can rearrange the vector $(\psi_0,\psi_1,\psi_2,\psi_3,\cdots)$ into $(\psi_1,\psi_3,\cdots,\psi_0,\psi_2,\psi_4,\cdots)$. In this basis, the aforementioned matrices can be redefined as follows:

For the two-site lattice:
\begin{equation}
C_3'=\frac{1}{4}\left ( \begin{array}{ccc}
            1      &   0           & 0      \\
           0       &  4h_L^2       &  2h_L  \\
           0       &  2h_L        &  1+4h_R^2 
         \end{array} \right ).
\end{equation}

For the four-site lattice:
\begin{equation}
C_5'=\frac{1}{4}\left ( \begin{array}{ccccc}
           1  & 1    &  0        & 0           &   0    \\
           1  & 2    &  0        & 0           &   0    \\
           0  & 0    & 4h_L^2    & 2h_L      &  0      \\
           0  & 0    &  2h_L     & 2         &   1    \\
           0  & 0    &0         & 1         &   1+4h_R^2   \\
         \end{array} \right ).
\end{equation}

And for the six-site lattice:
\begin{equation}
C_7'=\frac{1}{4}\left ( \begin{array}{ccccccc}
         1   & 1   & 0    & 0         & 0       & 0  & 0  \\
         1   & 2   & 1    & 0         & 0       & 0  & 0  \\
         0   & 1   & 2    & 0         & 0       & 0  & 0  \\
         0   & 0   & 0    &  4h_L^2   &  2h_L   & 0  & 0  \\
         0   & 0   & 0    &  2h_L     &   2     & 1  & 0\\
         0   & 0   & 0    & 0         & 1       & 2  & 1 \\
         0   & 0   & 0    & 0         & 0       & 1  & 1+4h_R^2
         \end{array} \right ).
\end{equation}

To solve the eigenvalues, we introduce the matrix $D = 4(C' - \varepsilon^2 I)$, where $I$ represents the unit matrix. The eigenvalues can be obtained by solving the equation $|D| = 0$, where $|D|$ denotes the determinant of matrix $D$. Specifically, for $C_3'$, the matrix $D$ is given by:

\begin{equation}
D_3=4(C_3' - \varepsilon^2 I_3)=\left ( \begin{array}{cc}
           E_1     & 0  \\
          0  &   F_2 
          \end{array} \right ),
\end{equation}

where

\begin{equation}
E_1=1-4\varepsilon^2=x-1, 
\label{eq:E1}
\end{equation}
and
\begin{eqnarray}
F_2 & = & \left ( \begin{array}{cc}
            4h_L^2 - 4\varepsilon^2   & 2h_L^2    \\
           2h_L^2      & 1+4h_R^2 - 4\varepsilon^2  
          \end{array} \right ) \nonumber \\
  & =  &\left ( \begin{array}{cc}
            L + x - 1    & \sqrt{L + 1}    \\
           \sqrt{L + 1}      & R + x  
          \end{array} \right ),
\label{eq:F2}
\end{eqnarray}
Here the variables are defined as
\begin{equation}
x=2-4\varepsilon^2,  ~~~L=4h_L^2-1,~~~R=4h_R^2-1
\label{eq:rl}
\end{equation}

The matrix $D$ corresponding to $C_5'$ is given by
\begin{equation}
D_5=\left ( \begin{array}{cc}
           E_2     & 0  \\
          0  &   F_3 
          \end{array} \right ).
\end{equation}
where 
\begin{equation}
E_2=\left ( \begin{array}{cc}
           x-1  & 1     \\
           1  & x   
          \end{array} \right ),
\label{eq:E2}
\end{equation}
and
\begin{equation}
F_3=\left ( \begin{array}{ccc}
           L+x-1           & \sqrt{L+1}  &  0  \\
           \sqrt{L+1}      & x           &  1     \\
             0             & 1           &   R+x   
          \end{array} \right ).
\label{eq:F3}
\end{equation}

The matrix $D$ corresponding to $C_7'$ is given by
\begin{equation}
D_7=\left ( \begin{array}{cc}
           E_3     & 0  \\
          0  &   F_4 
          \end{array} \right ).
\end{equation}
where 
\begin{equation}
E_3=\left ( \begin{array}{ccc}
           x-1  & 1 & 0    \\
           1  & x & 1    \\
           0  & 1 & x    
          \end{array} \right ), 
\label{eq:E3}
\end{equation}
and
\begin{equation}
F_4=\left ( \begin{array}{cccc}
            L+x-1    & \sqrt{L+1}      &  0    &  0    \\
            \sqrt{L+1}     & x          &  1    &  0   \\
            0        &1           &  x    & 1    \\
            0        &0            & 1    & R+x   
          \end{array} \right ).
\label{eq:F4}
\end{equation}
Generally we have
\begin{equation}
D_{2n+1}=\left ( \begin{array}{cc}
           E_n     & 0  \\
          0  &   F_{n+1} 
          \end{array} \right ).
\end{equation}
where $E_{n}$ and $F_{n+1}$ can be obtained by extending the above discussion.
It has
\begin{equation}
|D_{2n+1}|=|E_{n}||F_{n+1}|.
\label{eq:def}
\end{equation}

The determinant of $E_{n}$ for $n=1,2,3$ can be calculated easily. From Eq. (\ref{eq:E1}),(\ref{eq:E2}) and (\ref{eq:E3}), we get
\begin{equation}
|E_1|=x-1,~~|E_2|=x|E_1|-1, ~~|E_3|=x|E_2|-|E_1|
\end{equation}
We notice that for $n>3$ the elements of $E_{n+1}$ are the same as $E_n$ in the first $n$ columns and $n$ rows, i.e. $(E_{n+1})_{i,j}=(E_n)_{i,j}$ for $i,j \le n$. The rest elements are given by $(E_{n+1})_{n+1,j}=(E_{n+1})_{j,n+1}=0$ for $j<n$, $(E_{n+1})_{n,n+1}=(E_{n+1})_{n+1,n}=1$, $(E_{n+1})_{n+1,n+1}=x$.
Therefore, we have
\begin{equation}
|E_{n+1}|=x|E_n|-|E_{n-1}|
\label{eq:Eiteration}
\end{equation}

The calculation of  $|F_{n+1}|$ for $n=1,2,3$  is a bit tricky. From Eq. (\ref{eq:F2}),(\ref{eq:F3}) and (\ref{eq:F4}), we get
\begin{eqnarray}
|F_2| & = & (x+L+R)|E_1|+ RL-1 \nonumber \\
|F_3| & = & (x+L+R)|E_2|+ (RL-1)|E_1| \nonumber \\
|F_4| & = & (x+L+R)|E_3|+ (RL-1)|E_2|
\end{eqnarray}
It can be expected that for $n > 3$
\begin{equation}
|F_{n+1}|=(x+L+R)|E_n|+ (RL-1)|E_{n-1}|
\end{equation}
Now we prove this relation. From Eq. (\ref{eq:F3}) we get
\begin{equation}
|F_3|=(R+x)|G_2|-|G_1|
\end{equation}
where
\begin{eqnarray}
G_1 & = & (L+x-1) = ( L+|E_1| ), \nonumber \\
G_2 & = & \left ( \begin{array}{cc}
            L+x-1    & \sqrt{L+1}   \\
            \sqrt{L+1}     & x    
          \end{array} \right ).
\end{eqnarray}
It has 
\begin{equation}
|G_2|=|E_2|+L|E_1|
\end{equation}

From Eq. (\ref{eq:F4}), we get
\begin{equation}
|F_4|=(R+x)|G_3|-|G_2|
\end{equation}
where
\begin{equation}
G_3 = \left ( \begin{array}{ccc}
            L+x-1          & \sqrt{L+1} & 0  \\
            \sqrt{L+1}     & x          & 1 \\
             0             & 1          & x 
          \end{array} \right ).
\end{equation}
Obviously
\begin{eqnarray}
|G_3| & = & x|G_2|-|G_1| \nonumber \\
      & = & x|E_2|-|E_1|+L(x|E_1|-1) \nonumber \\
      & = & |E_3|+L|E_2|
\end{eqnarray}

We notice that for $n>2$ the elements of $G_{n+1}$ are the same as those of $G_n$ in the first $n$ columns and $n$ rows, i.e. $(G_{n+1})_{i,j}=(G_n)_{i,j}$ for $i,j \le n$. The rest elements are given by $(G_{n+1})_{n+1,j}=(G_{n+1})_{j,n+1}=0$ for $j<n$, $(G_{n+1})_{n,n+1}=(G_{n+1})_{n+1,n}=1$, $(G_{n+1})_{n+1,n+1}=x$.
Therefore, we have
\begin{equation}
|G_{n+1}|=x|G_n|-|G_{n-1}|.
\label{eq:Giteration}
\end{equation}
If
\begin{equation}
|G_n|=|E_n+L|E_{n-1}|
\end{equation}
it leads to
\begin{equation}
|G_{n+1}|=|E_{n+1}|+L|E_n|.
\label{eq:gn}
\end{equation}
The iteration rule Eq. (\ref{eq:Eiteration}) is used.

Therefore we have
\begin{eqnarray}
|F_{n+1}| & = & (R+x)|G_n|-|G_{n-1}| \nonumber \\
         & = & (x+L+R)|E_n|+ (RL-1)|E_{n-1}|
\end{eqnarray}
The iteration rule Eq. (\ref{eq:Eiteration}) is used.

From Eq. (\ref{eq:def}), we get
\begin{equation}
|D_{2n+1}|=|E_n|((x+L+R)|E_n|+ (RL-1)|E_{n-1}|)
\end{equation}
If
\begin{equation}
RL-1=0,
\end{equation}
the roots in $D_{2n+1}=(x+L+R)|E_n|^2=0$ are doubly degenerated.Therefore the ground state is degenerated doubly, the phase transition occurs. From Eq. (\ref{eq:rl}), the condition $RL-1=0$ just gives the phase boundary
\begin{equation}
\frac{1}{h_L^2}+\frac{1}{h_R^2}=4.
\end{equation}

\section{The phase boundary for the lattice with odd sites}
To be convenient to the readers, we show the matrix $C$ for the model with three sites, five sites and seven sites. Since odd sites only couple with odd sites and even sites only couple with even sites, we can rearrange the vector $(\psi_0,\psi_1,\psi_2,\psi_3,\cdots)$ into $(\psi_1,\psi_3,\cdots,\psi_0,\psi_2,\psi_4,\cdots)$. In this kind of bases, for the model with three sites, the Matrix $C'$ is given by
\begin{equation}
C_4'=\frac{1}{4}\left ( \begin{array}{cccc}
            1      &   1           & 0     & 0     \\
            1      &   1+4h_R^2    & 0     & 0     \\
            0      &       0       &  4h_L^2      &  2h_L  \\
            0      & 0             &  2h_L        &  2 
         \end{array} \right ).
\end{equation}

For the lattice with four sites, it is given by
\begin{equation}
C_6'=\frac{1}{4}\left ( \begin{array}{cccccc}
           1  & 1    &  0        & 0           &   0   & 0  \\
           1  & 2    &  1        & 0           &   0   & 0  \\
           0  & 1    & 1+4h_R^2  & 0           & 0     & 0   \\
           0  & 0    & 0         & 4h_L^2      & 2h_L      &  0      \\
           0  & 0    & 0         &  2h_L       & 2         &  1    \\
           0  & 0    &0          & 0           & 1         &  2   \\
         \end{array} \right )
\end{equation}
For six sites lattice it is given by
\begin{equation}
C_8'=\frac{1}{4}\left ( \begin{array}{cccccccc}
         1   & 1   & 0    & 0         & 0         & 0       & 0  & 0 \\
         1   & 2   & 1    & 0         & 0         & 0       & 0  & 0 \\
         0   & 1   & 2    & 1         & 0         & 0       & 0  & 0 \\
         0   & 0   &1     & 1+4h_R^2  & 0         & 0       & 0  & 0  \\
         0   & 0   & 0    & 0         &  4h_L^2   &  2h_L   & 0  & 0  \\
         0   & 0   & 0    & 0         &  2h_L     &   2     & 1  & 0\\
         0   & 0   & 0    & 0         & 0         & 1       & 2  & 1 \\
         0   & 0   & 0    & 0         & 0         & 0       & 1  &  2
         \end{array} \right )
\end{equation}
Similarly we introduce the matirx $D=4(C-\varepsilon^2 I)$ where $I$ is the unit matrix. The matrix $D$ for the above three matrixes can be written by
\begin{equation}
D_{2n}'=\left ( \begin{array}{cc}
            G'_n      &   0     \\
            0         &   G_n     
         \end{array} \right ).
\label{eq:gg}
\end{equation}
where $G_n$ is discussed around Eq. (\ref{eq:Giteration}) and $G_n'$ for $n=2,3,4$ are given by
\begin{equation}
G_2'=\left ( \begin{array}{ccc}
           x-1  & 1    \\
           1    & R+x    
         \end{array} \right )
\end{equation}
\begin{equation}
G_3'=\left ( \begin{array}{ccc}
           x-1  & 1    &  0         \\
           1  & x    &  1         \\
           0  & 1    & R+x    
         \end{array} \right )
\end{equation}
\begin{equation}
G_4'=\left ( \begin{array}{cccc}
           x-1  & 1    &  0    &0        \\
           1    & x    &  1   &0        \\
           0    & 1    & x     &  1    \\
           0    & 0    & 1    & R+x    
         \end{array} \right )
\end{equation}
We notice that the elements of $G_n'$ are the same as $E_{n-1}$ in the first $n-1$ rows and columns, so we have
\begin{eqnarray}
|G_n'| & = & (R+x)|E_{n-1}|-|E_{n-2}| \nonumber \\
       & = & |E_n|+R|E_{n-1}|.
\label{eq:gn1}
\end{eqnarray}
The iteration rule Eq. (\ref{eq:Eiteration}) is used. 

From Eq. (\ref{eq:gg}), (\ref{eq:gn}) and (\ref{eq:gn1}), we get the determinant of $D_{2n}$
\begin{equation}
|D_{2n}|=(|E_n|+R|E_{n-1}|)(|E_{n+1}|+L|E_n|)
\end{equation}
If
\begin{equation}
R=L,
\end{equation}
the roots of $|D_{2n}|=0$ are doubly degenerated. Hence the ground state is doubly degenerated and the phase transition occurs. The condition $R=L$ just gives the phase boundary
\begin{equation}
h_L=h_R
\end{equation}


\begin{thebibliography}{}
\bibitem{hami} The simplest example is given by $H=\left ( \begin{array}{cc}
           -1  & 0        \\
           0    & \lambda    
         \end{array} \right )=h_L+h_R(\lambda)$, where $h_L=\left ( \begin{array}{cc}
           -1  & 0        \\
           0    & 0    
         \end{array} \right )=h_L+h_R(\lambda)$ and $h_R(\lambda)=\left ( \begin{array}{cc}
           0  & 0        \\
           0    & \lambda    
         \end{array} \right )=h_L+h_R(\lambda)$. For $\lambda<-1$, the ground state energy is $\lambda$ and for $\lambda>-1$, it is $-1$.
\end{thebibliography}
\end{document}